\documentclass[%
 reprint,
superscriptaddress,
 amsmath,amssymb,aps,pra,
]{revtex4-2}
\usepackage{color}
\usepackage{graphicx}
\usepackage{dcolumn}
\usepackage{bm}
\usepackage{epstopdf} 
\usepackage{hyperref}
\usepackage{textcomp}
\usepackage{braket}

\begin{document}

\preprint{APS/123-QED}

\title{Stabilization of ion-trap secular frequencies for a quantum phase transition study}

\author{Jie Zhang}
\affiliation{ Department of Physics, Simon Fraser University, Burnaby, BC, Canada, V5A 1S6}
\affiliation{ Institute for Quantum Science and Technology, National University of Defense Technology, Changsha, Hunan, China, 410073}
\author{B.~T. Chow}
\affiliation{ Department of Physics, Simon Fraser University, Burnaby, BC, Canada, V5A 1S6}
\author{ P.~C. Haljan}
\email{phaljan@sfu.ca}
\affiliation{ Department of Physics, Simon Fraser University, Burnaby, BC, Canada, V5A 1S6}

\date{\today}

\begin{abstract}
An array of ions in a linear radio-frequency (RF) Paul trap is a good candidate for investigating structural phase transitions, such as the linear-to-zigzag (LZ) transition, due to the convenient control provided by modification of the trap confinement strength. In such studies, the trap secular frequencies are a key factor that limits the stability of the critical point (CP). In this paper, we implement secular-frequency stabilization, including active feedback stabilization of the RF voltage near the trap electrodes, and achieve a stability of better than 5~ppm over 200~s for both transverse and axial potentials. To evaluate the combined long-term stability of the trap potential in both directions, we measure the zigzag (ZZ) mode frequency near the CP, where the effect of instability in both trap directions is substantially amplified. The short-term noise within a limited spectral window is also suppressed by this scheme, as inferred from decoherence rates of the ZZ mode measured at different frequencies with Ramsey fringes. We also identify that the limitation of the secular frequency stability is mainly from the imperfect temperature insensitivity of voltage dividers and RF detectors, and as a result improvement of lab temperature stability is of great help for stabilizing the trap frequency.
%
%
\end{abstract}

\maketitle


\section{Introduction}

In a linear radio frequency  (RF) Paul trap, a linear string of ions undergoes a structural transition to a zigzag ion-crystal configuration when the ratio of transverse and axial secular oscillation frequency is reduced across the transition CP \cite{pyka2013topological,Ulm2013KZM,Sara2013Kink}. When the ion string is Doppler laser-cooled, the thermal excitation energy results in classical dynamics of the LZ transition, which can be controlled to nucleate topological kinks and perform studies of the Kibble-Zurek Mechanism \cite{pyka2013topological,Ulm2013KZM,Sara2013Kink}. With additional laser cooling technology, the ion string can be further cooled to the motional ground state by electromagnetically-induced-transparency ground-state cooling \cite{PhysRevLett.85.4458, PhysRevLett.85.5547, PhysRevA.93.053401, qiao2020doubleeit, PhysRevApplied.18.014022,PhysRevLett.125.053001} or sideband cooling \cite{monroe1995resolved, wineland1998twoions,IonQ2018PRA}, possibly in conjunction with Sisyphus cooling \cite{ejtemaee20173d,joshi2020polarization}. Ultra-cold ion strings then open a path for studying the dynamics of LZ transition  in the quantum regime \cite{retzker2008double, zhang2022spectroscopic}. Near the LZ transition CP of the ion string, the ZZ mode frequency has very high sensitivity to changes in the confinement strength of the ion trap. While this can potentially be used to probe noise in the trap potential ranging from the DC to MHz, the sensitivity implies that studying quantum properties near the CP of the LZ transition requires stringent stability of the secular trap frequencies. For other applications such as quantum information processing \cite{blatt2008entangled} and quantum simulation \cite{Monroe2014Nature,BlattEntangleManyBody2014},  atomic interferometry \cite{johnson2015sensing}, and quantum metrology \cite{chou2010frequency}, stable trap secular frequencies also play an important role.

Ions are confined by a combination of RF and DC electrical potentials in a linear Paul trap, where the RF source typically works at the frequency of several tens of MHz and is generally amplified and then coupled to an resonant effective-LC circuit consisting of a helical RF resonator attached to the ion trap. In this setup the RF voltage amplitude at the trap electrodes can be up to several hundred volts. Since there are multiple components in the RF circuit, any electrical fluctuation in the RF source and amplifier, or temperature induced fluctuations in the electronic components, will cause a change in the trap secular frequencies. For our purpose of investigating quantum dynamics near the LZ transition of ion string, we need highly stabilized trap secular frequencies; therefore, an active feedback loop should be employed to suppress this noise and stabilize the trap.

For the  stabilization of the transverse secular frequency, we capacitively sample the RF signal near the vacuum feedthrough to the ion trap and feed back to the RF source to stabilize the voltage at the RF electrodes \cite{johnson2016active}.   The RF source is non-invasively sampled by two temperature insensitive voltage dividers from the end of RF coils of a double-coil helical resonator and the sampled RF signals are sent to two separated temperature compensated RF rectifiers,  which rectify the RF signals to DC signals. One of the DC signals goes to the servo loop for locking the RF potential, while the other is monitored by a digital multimeter to identify the stability of RF system. The axial trap secular frequency shows decent passsive stability since the DC voltage system is robust to ambient noise and temperature fluctuations. The long-term stability of trap secular frequencies in both axial and transverse directions is assessed using a single trapped $^{171}$Yb$^{+}$ ion and the combined stability is evaluated by monitoring the stability of a 4-ion ZZ mode near the CP, where the effect on the mode frequency of instabilities in both trap directions is amplified. The short-term stability of the trap potential is assessed from the decoherence rate of the ZZ mode, which is measured with a Ramsey fringe technique. We also found that the long-term drift performance of the trap secular frequencies highly depends on the ambient temperature stability, which can be improved by using a better environmental control system.

The structure of this paper is organized as follows: In Sec.~\ref{sec:theory}, we give the background of secular trap frequency stabilization for the study of the LZ structural phase transition and provide a model of the linear ion trap potential. The experimental setup of our trapped ion system is introduced in Sec.~\ref{sec:setup}. The experimental results for long- and short-term stability are presented in Sec.~\ref{sec:results}. Finally, the conclusions of our paper are given.

\section{Linear ion trap theory}\label{sec:theory}

\begin{figure}
	\centering
	\includegraphics[width=0.8\linewidth]{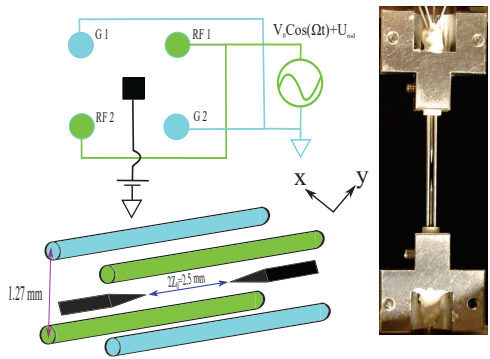}
	\caption{The linear  Paul trap used in our experiment is composed of four rods held in a square configuration of the side of 1.27 mm and two needles separated by 2.5 mm. The rods confine the ions in the transverse direction, two opposing rods are grounded (shown in blue) and the other two are connected to an rf source of $V_0\cos(\Omega t)+U_{rod}$ (shown in olive green). Two needles confining the ions in the axial direction  are connected to a DC voltages of $U_0$.}
	\label{fig:figure2_1}
\end{figure}

In our experiment, the ions are confined in a linear Paul trap, which consists of four tungsten rods and two tungsten end-cap needles, see Fig. \ref{fig:figure2_1} for details.
 The four rods have a diameter of 0.48 mm with a rod to rod  distance of 1.27 mm and  the end-cap needles have a space of $2z_0 = 2.5$ mm from tip to tip. RF potential applied to one of the two pairs of diagonal electrodes and ground connected to the other pair provide the radial confinement. In the axial direction, A dc voltage applied to the end-cap creates the confinement along $z$-axis. Recently a detailed numerical simulations of our linear trap \cite{MattFrick2016} give the precise electric potential near the trap axis:
\begin{equation}\label{eqn1}
\Phi \approx (\dfrac{V_0\cos(\Omega_Tt) + U_{rod}}{2})[const + C_{22}\dfrac{x^2-y^2}{R_0^2} + C_{20}\dfrac{r^2-2z^2}{R_0^2}]
\end{equation}
where $V_0$ is the RF amplitude, $U_{rod}$ is the DC voltage applied to the radial electrodes, $\Omega_T$ is the RF frequency and the $R_0$ is the distance from the ion to the rod surface. $C_{22}$ and $C_{20}$ are the trap geometrical parameters and have been determined numerically to be
\begin{equation}\label{eqn2}
C_{22}=0.889; {C}_{20}=0.0106\ll {C}_{22}
\end{equation}
In Eqn. \ref{eqn1}, $C_{22}$ would be unity for an ideal transverse quadrupole potential while its actual value in Eqn. \ref{eqn1}  leads to the effective radius $R_0$ introduced above. The third term in Eqn. \ref{eqn1} is responsible for the transverse fields (both RF and DC on the rods) leaking into the axial direction and thereby affecting the ion motion in the $z$-axis.

A single ion with charge $e$ and mass $m$ oscillating in the Paul trap has a harmonic oscillation frequency \cite{Wolfgang1990}
\begin{equation}\label{eqn3}
\omega_r = \dfrac{eV_0}{\sqrt{2}m\Omega_T R^2}
\end{equation}
in the radial direction. Here we have assumed the $\omega\ll\Omega_T$ and ignored the negligible residual transverse forces from the static potentials.  While the axial motional  harmonic oscillation frequency
\begin{equation}\label{eqn4}
\omega_z = \dfrac{\sqrt{2\kappa eU_0}}{\sqrt{m}Z_0}
\end{equation}
is determined by the static potential $U_0$ on the endcap tip, the distance from  trap center to the end-cap tip and geometry factor $\kappa$.

To stabilize the secular trap frequency in the transverse direction, the RF amplitude $V_0$ and RF frequency $\Omega_T$ should be stabilized since other parameters are robust to the existing noise.  The resonant RF frequency is determined by the conductance of the quarter-wave resonator and the capacitance of the feedthrough and trap itself. Since RF source is provided by a digital RF generator, which has stability more than we desire, we just need to focus on stabilizing the RF amplitude $V_0$. For the axial secular trap frequency, which dependents only on the static potential amplitude, stable DC voltage should guarantee the required stability.

A string of ions stored in the trap stays linear structure as long as the radial confinement is considerably stronger than the axial confinement. The linear ion string  will experience a structural phase transition from the linear to zigzag when damping the radial potential (y direction in our experiment) across the transition CP, which is  indicated by the
\begin{figure}
	\centering
	\includegraphics[width=0.8\linewidth]{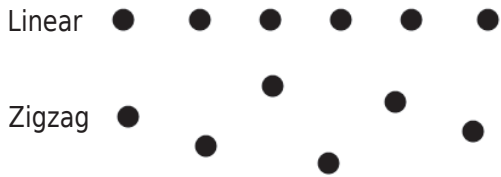}
	\caption{Structure of 6 ions in the linear Paul trap. When the anisotropy is 6.8, the transverse confinement is much stronger than the axial confinement, the ion string stays linear structure. Damping the transverse confinement with constant axial confinement, which reduces the anisotropy to 3.4, the ion undergoes a structural phase transition from the linear to zigzag. }
	\label{fig:figure2_2}
\end{figure}
 anisotropy at $\alpha_c=\omega_{yc}/\omega_z$, where $\omega_{yc}$ is the secular frequency of center of mass (COM) mode at the CP and $\omega_z$ is the axial COM mode frequency, which is usually kept constant in the structual phase transition experiment. When the anisotropy is less than  $\alpha_c$, a structural phase transition occurs and the linear ion structure turns into a 2-D zigzag structure, see Fig. \ref{fig:figure2_2} for an example. In the process of damping the radial confiment, the zigzag mode frequency ($\omega_{yzz}$) of the ion string follows \cite{Jixin2012}
\begin{equation}\label{eqn5}
\omega_{yzz} = \sqrt{\omega_{y}^2-\omega_{yc}^2}
\end{equation}
on the linear side.

 Eqn. (\ref{eqn5}) indicates that the  zigzag mode frequency is determined by  the current COM mode frequency $\omega_y$ and the  COM mode frequency $\omega_{yc}$ at  CP, where the ZZ mode vanishes ($\omega_{yzz}=0$). Therefore any drift from the transverse or axial COM mode would cause the ZZ mode drift. We also find that the drift of the ZZ mode $\delta \omega_{yzz}$ is a factor of $\omega_{y}/\omega_{yzz}$ bigger than the drift of COM mode $\omega_y$, this indicates that we can measure the long term ZZ mode frequency stability to assess the   axial and transverse COM mode frequency stability at the same time with high sensitivity. The short term stability of ZZ mode is also crutial for investigating the coherence near CP, which can be assessed by using Ramsey fringes.

\section{Experimental setup}\label{sec:setup}

\begin{figure}
	\centering
	\includegraphics[width=0.95\linewidth]{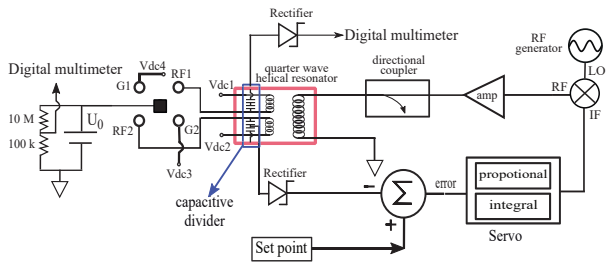}
	\caption{Trap frequency stabilization setup of the ion trap system. The output of RF frequency from the generator is fixed as the local oscillator for the mixer, the IF port of the mixer is connected to the feedback loop which regulates the output of the RF port. The RF amplitude is sampled two by voltage dividers and rectifiers, while the DC voltage is sampled by using two resistors.   }
	\label{fig:figure3_1}
\end{figure}

Stabilization of transverse potential is similar to the method usd in Ref. \cite{johnson2016active}. The high voltage RF signal is  sampled  at the end of  a double-coil quarter-wave helical resonator and sent to a servo which regulates the input RF amplitude. The detailed setup of the locking scheme is shown in Fig. \ref{fig:figure3_1}.
RF  generator (Keysight 34455B) produces an rf signal at $\Omega/2\pi = 16.9 $ MHz and -6.7 dBm for the local oscillator port of a level 3 frequency mixer (Mini-Circuits ZX05-1L-S). The RF port of the mixer  is connected to an RF amplifier that amplifies the RF signal by  30 dB. The amplified signal inductively couples to the double-coil quarter-wave helical resonator through a primary coil, which provides impedance matching between the rf source and the circuit formed by the resonator and ion trap electrode. The loaded Q factor for the RF system is about 180. The double-coil resonator provides separate connection for each RF electrode that allows the DC voltage to be applied to compensate the micromotion and tune the secular frequency \cite{SaraPhD2015} of ions.

The noninvasive sampling of the   RF voltage from the system is realized by using same capacitive voltage dividers and temperature compensated rectifiers as described in ref. \cite{johnson2016active}. Due to the capacitance of the circuit board mounting the low temperature coefficient capacitors, the voltage divider samples about 2\% of the input RF amplitude rather than the designed 1\%. Considering the circuit consisting  of divider and rectifier,  the ratio of the rectified dc output voltage to rf input voltage amplitude is about 1:250 at a drive frequency of 16.9 MHz. We also construct a stable setpoint box using Linear Technology LTC6655 5V reference mounted on a DC2095A-C evaluation board, the setpoint voltage can be varied by  Analog Devices EVAL-AD5791 and ADSP-BF527 interface board with 20-bit precision and 0.25 ppm stability. The difference between these inputs, i.e., the error signal, is then sent to the servo box (New Focus LB1005) which generates the feedback signal to IF port of the mixer and  stabilizes the RF amplitude at the  RF sampling point. As we can see in the Fig. \ref{fig:figure3_1} there are two dividers for sampling the the RF voltage, only one of them is used to feedback the RF circuit and the other one is  connected to the digital multimeter (DMM) for monitoring the stability of RF voltage on the rod.  Due to the  DC voltage dependence of axial trap frequency, we just need to  monitor the stability of end-cap DC voltage. In our setup a resistive voltage divider, composing of  10 M$\Omega$ and 100 $k\Omega$ resistors, is constructed to sample about 1\% of the end-cap voltage.

We  characterize the long term stability of the locked RF system and DC potential by measuring the  motional oscillation frequency of a single $^{171}$Yb$^+$ ion in both transverse and axial directions.  We also measure the stability of ZZ mode frequency of 4 ion-string on linear side as it is a sensitive indicator for the combined stability of both transverse and axial COM modes. The qubit state $\ket{F=0, m_F=0} \equiv\ket{\downarrow}$ and  $\ket{F=1, m_f=0} \equiv \ket{\uparrow}$, which are the hyperfine levels of the $^2S_{1/2}$ of $^{171}$Yb$^+$ with energy splitting $\omega_{HF}/2\pi=$12.642815 GHz, is measured by microwave transition to identify the stability of the qubit itself. Motional sensitive two-photon Raman spectroscopy \cite{Ramsey1990} is employed to acquire the motional sideband $\omega_{HF}\pm\omega$ with transverse harmonic oscillation frequency $\omega_{r}/2\pi\approx 0.85 $ MHz and axial motional frequency $\omega_{z}/2\pi\approx 0.325 $ MHz at end-cap voltage of 78 V.

The whole experimental process of single ion  includes several steps, first Doppler cooling and Sisyphus cooling method \cite{ejtemaee20173d} is used to cool the single ion  to a phonon number about 1.5 with laser wavelength at 369.5 nm. The ion is then prepared in the $\ket{\downarrow}$ state through optical pumping, and following the Raman spectroscopy is applied to detect the motional sensitive blue sideband transition (BSBT), finally the state ($\ket{\uparrow}$) is measured with state-dependent fluorescence techniques \cite{Olmschenk2007,ejtemaee2010}. For the ion-string test of ZZ mode frequency stability, the measuring process is almost same as the single ion test except that the resolved Raman sideband cooling method \cite{leibfried2003quantum} is used to cool the transverse ZZ modes of ion-string to their motional ground states and the desired ZZ mode frequency is acquired by adiabatically ramping the DC voltage on the trap electrodes in quadrupole configuration.

\begin{figure}
	\centering
	\includegraphics[width=0.90\linewidth]{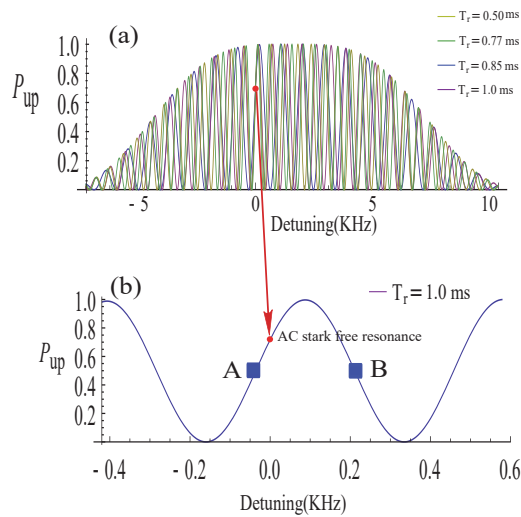}
	\caption{Raman spectroscopy  simulation for monitoring trap frequency drift. (a) Full Ramsey frequency scan simulation for BSBT at different Ramsey delay time, it is clear that the AC stark free resonance is independent of the Ramsey delay time, therefore we can exclude the ac Stark shift from our measurement. (b) Finer Ramsey frequency scan simulation for transverse BSBT near the AC stark free point, the curve has Ramsey time delay 1 ms. For fast sampling rate we choose to monitor the two points (point A and B) at the half fringe contrast near the central frequency and the sign of the probability difference at the two point indicates the trend of frequency drift.}
	\label{fig:figure3_3}
\end{figure}

Ramsey spectroscopy technique is used in our experiment to identify and monitor the motional BSBT at $\omega_{HF}+\omega$. The stability of the qubit detected by both microwave and optical Raman sideband transitions. The BSBTs of transverse and axial direction are driven by three Raman beams with two specific configurations that the can measure the transitions in the two directions separately. In our measurement, two $\pi/2$ pluses separated by a time delay $T_r=$ 1 ms drive the Raman transition in the transverse direction,  while the time delay is 10 ms for the axial direction for a precise measurement. After the Ramsey pulses, the probability of the $\ket{\uparrow}$ state is detected with 369.5 nm laser radiation. To reduce the fluctuation of the AC stark shift as much as possible, we monitor the Ramsey fringe closest to the point of ac Stark shift free resonance instead of the BSBT resonance as shown in Fig. \ref{fig:figure3_3}. The Ramsey fringe is fitted with empirical cosine function $ P_{\ket{\uparrow}}= Acos(\Delta T_r+ \phi) +$ offset, where $A$ is the contrast of the Ramsey fringes, $\Delta$ is the detuning between the Raman laser and BSBT resonance. The Ramsey frequency experiment is repeated 100 times for each value of $\Delta$ in order to acquire decent Ramsey fringes for extracting the $\omega$. As shown in Fig. \ref{fig:figure3_3}, we choose to monitor the two points (point A and B) with half contrast near the central frequency of the fringe for fast tracking of the  BSBT drift. In this method, the drift frequency can easily obtained from the unbalanced probability of point A and B.

\section{ Results: long-term trap potential stability}\label{sec:results}
\subsection{Verification of trap potential using a single ion}

\begin{figure}
	\centering
	\includegraphics[width=1.0\linewidth]{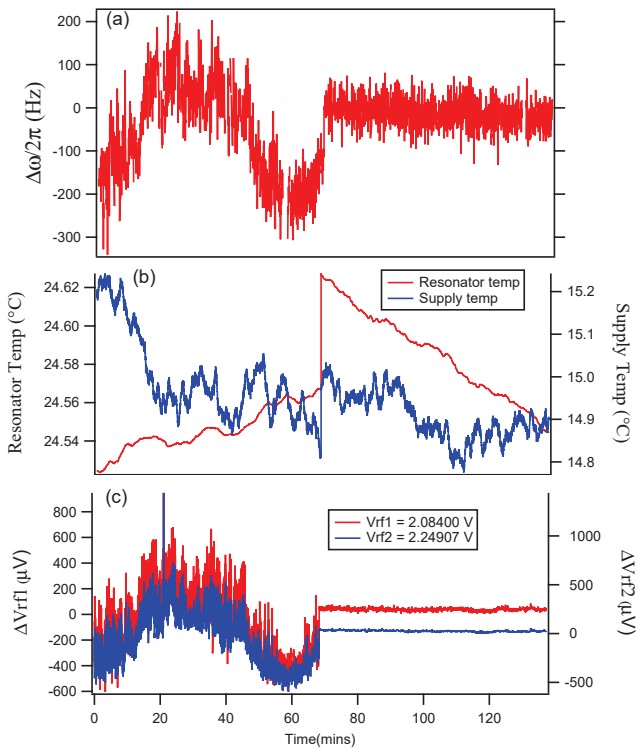}
	\caption{Transverse trap frequency, ambient temperature and sampled RF voltage drift monitored over the course of about 2 hours for both locked and unlocked cases. Note the locking started at about 70 mins after the data acquisition. (a) There is a clear suppression of the drift  with the locking scheme. (b) The resonator temperature  shows less than 0.2 $^{\circ}$C during the measurement time. (c) Sampled RF voltage with capacitive voltage divider and rectifier shows almost same variations with the trap frequency.}
	\label{fig:figure4-1}
\end{figure}

We monitored the qubit state and the transverse BSBT of a single ion for about 2 hours.  we can clearly see an improvement of the trap frequency stability  with the locking feedback on RF circuit as shown in Fig. \ref{fig:figure4-1}.(a). As a good indicator of the RF voltage amplitude, the sampled RF voltages labeled by Vrf1 and Vrf2 show same variations as the trap frequency in Fig. \ref{fig:figure4-1}.(c). The supply temperature near the vacuum chamber and on the surface of resonator are also monitored to acquire the correlations between the drift of trap frequency and ambient temperature. For stable temperature in the lab as shown in Fig. \ref{fig:figure4-1}.(b),  we can obtain more than 30 dB suppression of the long term drift for the transverse trap frequency.

\begin{figure}
	\centering
	\includegraphics[width=0.90\linewidth]{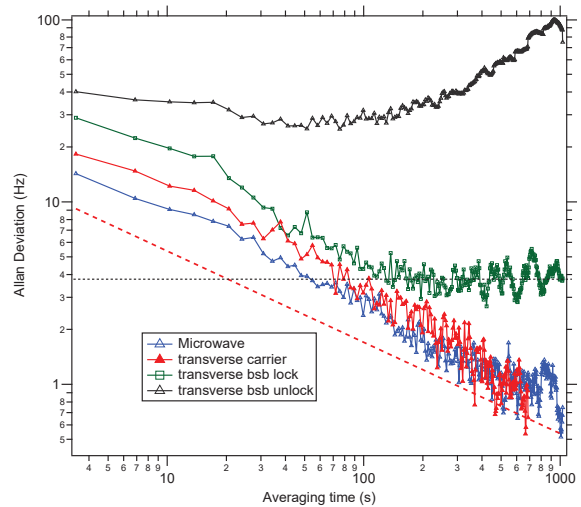}
	\caption{Allan deviations  of the transverse BSBT for the locked and unlocked cases, as well as the qubit carrier transition. The Allan deviation of the locked and unlocked BSBT are calculated from the  recorded data points shown in Fig. \ref{fig:figure4-1}(a). The carrier transition (qubit) curve is acquired from the measurement with both Raman transition and microwave transition.}
	\label{fig:figure4-2}
\end{figure}

We plot the Allan deviation \cite{Allan1966statistics} for the transverse trap frequency in the Fig. \ref{fig:figure4-2} as a function of the integration time $T$. The qubit state measurement is still limited by the white noise  as indicated by the dash line during the integration time up to 1000 s, while the BSBT in the locked case shows the an uncertainty after integration time about 200 s better than 4 Hz, i.e., 4.7 ppm. In the unlocked case, the BSBT deviation increases with the integration time, the stability of the trap frequency is affected by various of noise.

\begin{figure}
	\centering
	\includegraphics[width=1.0\linewidth]{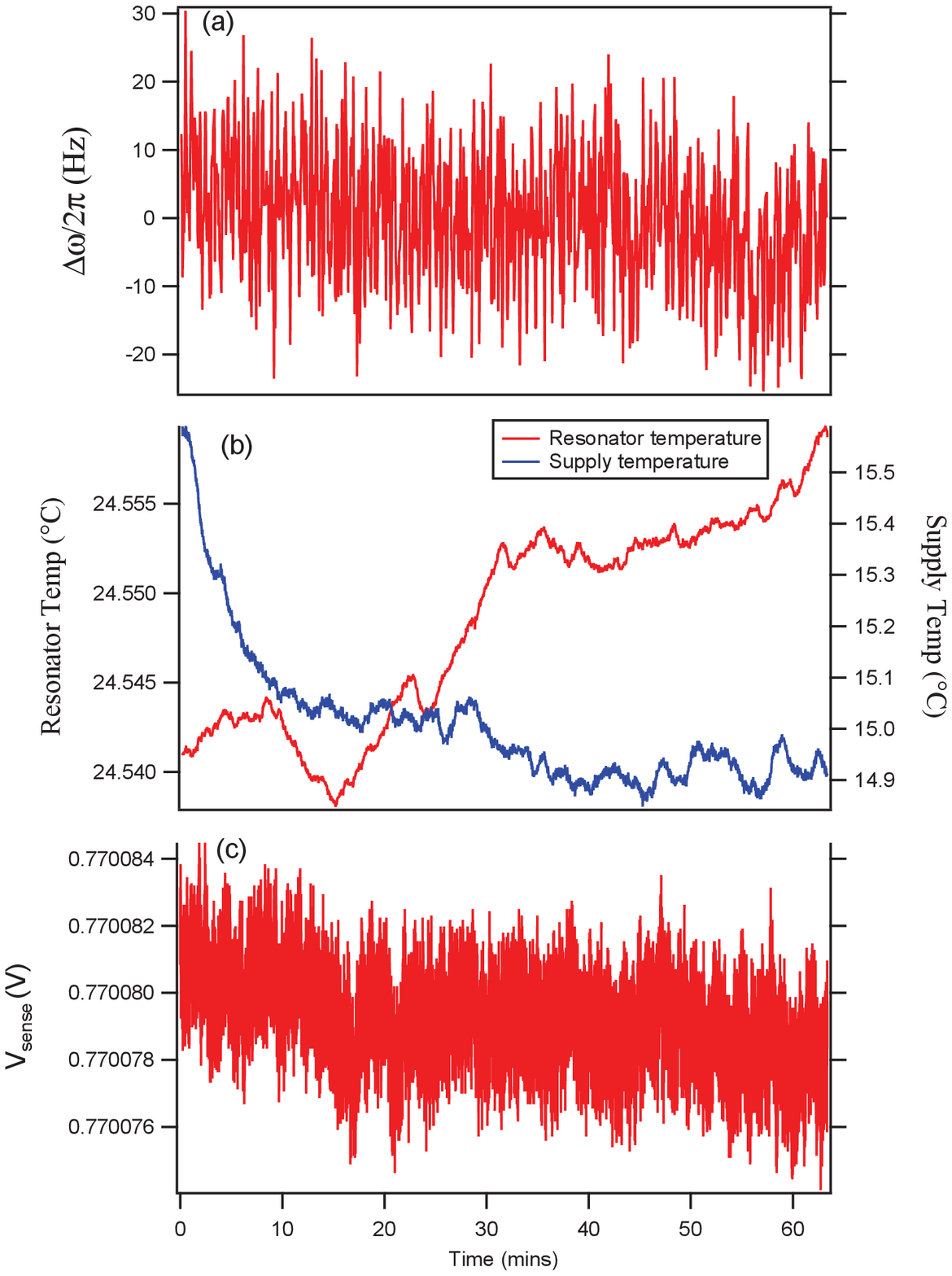}
	\caption{Axial trap frequency, ambient temperature and sampled end-cap voltage drift monitor for time of about 1 hour. Note the system is not locked.  (b) The ambient temperature in the lab shows about a fluctuation of 0.6 $^{\circ}$C during the measurement time. (c) The sampled end-cap voltage with resistive voltage divider shows less than 5 ppm variation.}
	\label{fig:figure4-3}
\end{figure}

The same method is also applied to measure the drift of axial trap frequency with end-cap voltage set at 78 V. The secular trap frequency, ambient temperature and sampled end-cap voltage  in the axial direction are monitored for time of about 1 hour as shown in Fig. \ref{fig:figure4-3}. The secular trap frequency shows high stability during the measurement as expected since the axial trap frequency is only determined by the end-cap voltage.

\begin{figure}
	\centering
	\includegraphics[width=0.850\linewidth]{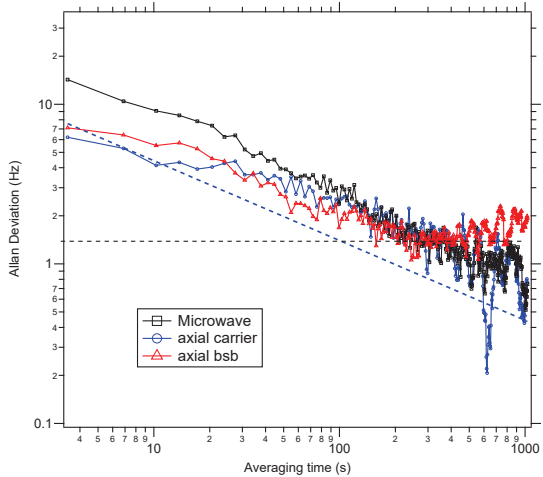}
	\caption{Allan deviations  of the axial secular frequency as well as the  carrier transition. The axial BSBT Allan deviation is calculated from the   recorded data points shown in Fig.\ref{fig:figure4-3}.(a), the carrier transition curve is acquired from a similar measurement by performing the Raman transition and microwave transition.}
	\label{fig:figure4-4}
\end{figure}

We also plot the Allan deviation for the axial frequency in the Fig. \ref{fig:figure4-4} as a function of the integration time $T$. The qubit state is similar to the case of transverse direction, which is limited by the white noise as indicated by the dash line for the integration time up to 1000 s. The BSBT in the locked case over an integration time about 300 s,  has an uncertainty of about 1.5 Hz, i.e., 4.6 ppm. Although we did not lock the end-cap voltage like the transverse direction, the stability of the axial trap frequency is still as good as the locked case in the transverse direction. Therefore there is no need to lock the end-cap voltage.

\begin{figure}
	\centering
	\includegraphics[width=1.0\linewidth]{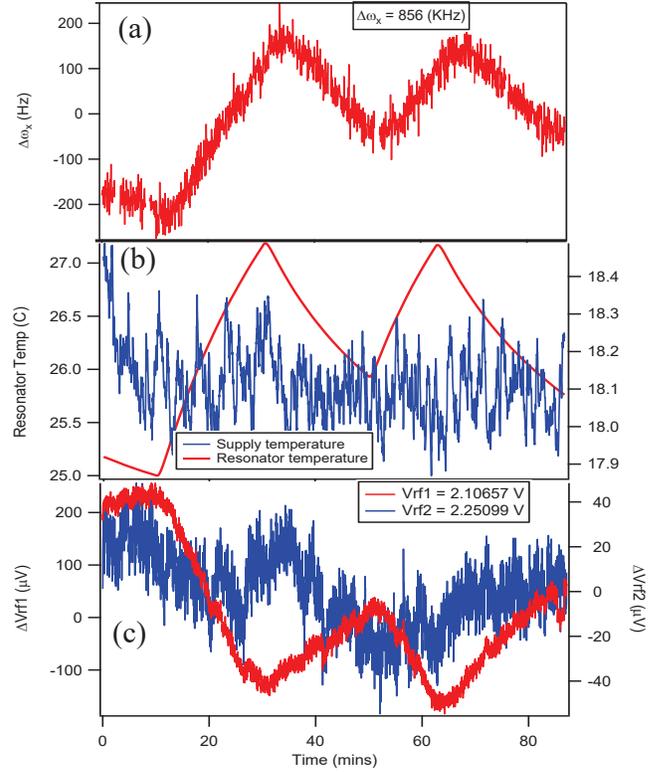}
	\caption{Temperature dependence of the transverse trap frequency and the sampled RF voltage. (a) Trap frequency varies with the resonator temperature. (b) Supply temperature stays stable while the resonator temperature is varied by using a heater. (c) Sampled RF voltage responses for  the variation of  resonator temperature. }
	\label{fig:figure4-5}
\end{figure}

\begin{figure}
	\centering
	\includegraphics[width=1.0\linewidth]{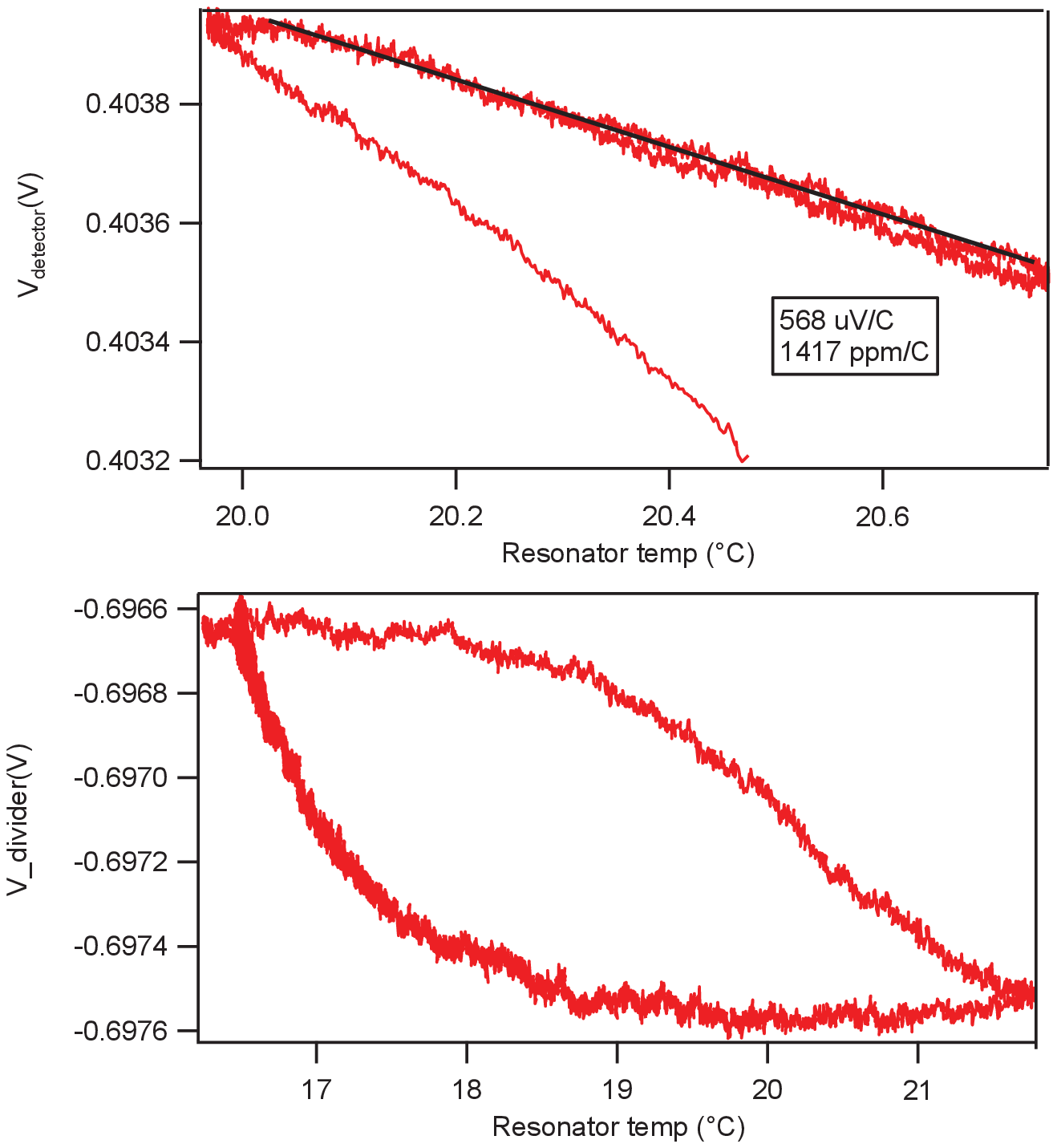}
	\caption{Temperature dependence tests for the RF detectors and capacitive voltage dividers. (a) Temperature dependence of the RF detector is tested by heating the detector directly. (b) Test  for the temperature dependence of the capacitive voltage divider is implemented by  heating the resonator. Note that the variations of the RF detector and voltage divider show different temperature dependence for increasing and decreasing temperature processes.  }
	\label{fig:figure4-6}
\end{figure}

By locking scheme in the transverse direction, we can get a stability better than 5 ppm, however there are still some factors limiting our scheme. One of the most important contributions is from the lab temperature, which puts affections on the components such as capacitive voltage dividers and RF rectifiers. In our experiment, the drift of the transverse trap frequency dependents heavily on the resonator temperature even in the locked case as shown in Fig. \ref{fig:figure4-5}. To demonstrate the correlation, we keep the RF system locked and heat the resonator, It can be seen that the system can not be locked stably with temperature varied.  We can easily find the correlation between the resonator temperature and variation of trap frequency. For a rough estimation, we need to achieve a resonator temperature stability about 0.3 $^{\circ}$C  to get  5 ppm of trap frequency  stability.

We further identify the voltage divider and rectifiers are the temperature sensitive part in the locking loop.  We build another similar RF system like the real one used for experiment to test the temperature sensitivity of the RF rectifier and capacitive voltage divider. The RF detector is tested by heating  itself directly, while we heat the resonator surface and build another rectifier placed far from the heating source to reliably measure the temperature dependence of the capacitive voltage divider. The temperature sensitivity for the RF detectors  is of 1400 ppm/ $^{\circ}$C as shown in the Fig. \ref{fig:figure4-6}. Note that the temperature dependence of the RF detector and voltage divider are different for the heating and cooling processes, we attribute the temperature sensors can not be close enough to the temperature sensitive parts.

\subsection{Long term trap potential stability near LZ transition}

\begin{figure}
	\centering
	\includegraphics[width=1.0\linewidth]{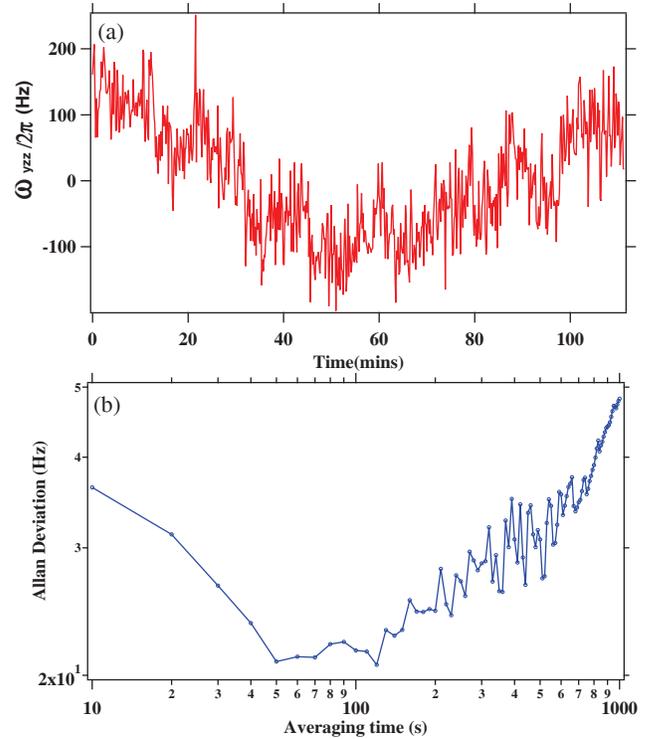}
	\caption{ZZ mode frequency stability measurement. (a) Data acquisition of ZZ mode at 56 KHz over a course of 2 hours. The ZZ mode frequency shows a less than 250 Hz drift.  (b)  Allan deviation of the ZZ mode at 56 kHz calculated from (a). The ZZ mode shows a stability of about 20 Hz for an integrating time of 100 seconds. }
	\label{fig:figure4-7}
\end{figure}

Quantum dynamics of LZ transition near critical point requires  us to assess the stability of trap potential since the CP is determined by trap frequency in both directions and the instability of trap frequency near the CP is significantly amplified. In our experiment, we choose to verify the potential stability by monitoring the ZZ mode frequency of  multiple ions string on the linear side. The ZZ mode frequency has higher sensitivity, which is amplified by a factor of the ratio of COM  and ZZ mode frequencies, than the COM mode.  Experimentally we adiabatically ramp the ZZ mode frequency of a 4-ion string from about 360 kHz to 56 kHz, where we use the fast trap frequency monitor method to record the ZZ mode frequency for over 2 hours. Fig. \ref{fig:figure4-7}(a) shows drift frequency of ZZ mode in 2 hours. The stability of the ZZ mode is worse  than the original COM mode as expected. Based on the data in Fig. \ref{fig:figure4-7}(a), we calculated the Allan deviation of ZZ mode in Fig. \ref{fig:figure4-7}(b) with a limit of about 20 Hz in an integration time of 100 s, which is clearly worse than either of the COM mode. 

\section{Results: short-term fluctuations}

In the experiment of investigating the quantum dynamics near the CP of  LZ transition, we not only require the long term stability of the ZZ mode, which has been verified in the previous section, but also the short term stability that limits the coherence time of the ZZ mode. To characterize the short term stability of ZZ mode frequency, we choose to monitor the decoherence rate of  ZZ mode of 4-ion string as function of different ZZ mode frequencies.  The fluctuations of ZZ mode frequency caused phase noise can be measured with  Ramsey experiment.  Since ZZ mode frequency can be varied ranging from 360 kHz to 0 Hz, we can also study ZZ mode frequency sensitivity to noise, i.e., use ion string as a probe for addressing the frequency dependent noise.

\begin{figure}
	\centering
	\includegraphics[width=1\linewidth]{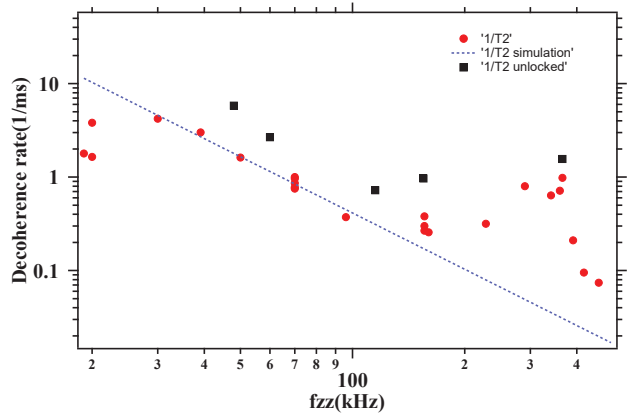}
	\caption{Ramsey decoherence rate of ZZ frequency on linear side with rf locked and unlocked. The red dots indicate the lock decoherence rate over a range of bout 400 kHz, while the black square denote the unlocked case with same ZZ mode frequency.  It is clear that the trap frequency stabilization improves the coherence of ZZ mode.  }
	\label{fig:figure5-1}
\end{figure}

As shown in Fig.\ref{fig:figure5-1}, we measure the decoherence rate of ZZ mode at different mode frequencies for both locked and unlocked cases. It can be seen that the decoherence rate of locked case is lower than unlocked case for almost the whole  ZZ mode frequency range, this results reveals that the servo also suppresses faster noise and improves coherence. However, the theoretical limit of  decoherence rate has not been reached in our experiment, this indicates that we still have technical noise to be figured out in our lock loop.

\section{Discussion and conclusion}

In this paper we have stabilized the transverse trap frequency and measured  less than 5 ppm long term stability for both transverse and axial directions.  We also verified the long  term  stability of the locking scheme by using the ZZ mode of a  4-ion string. We found the temperature sensitivity of the RF detector and capacitive voltage divider is the main reason that limits the long term stability of the trap frequency. Further improvement could be reached with better lab temperature control. Finally the fast frequency noise is also suppressed by our locking scheme, which is evaluated by using the decoherence rate of the ZZ mode of a 4-ion string.

\section{Acknowledgments}
This work was supported by NSERC and CFI LOF and by the China Scholarship Council and the National Natural Science Foundation of China (Grants No. 12004430)

\newpage
\nocite{*}

\bibliography{TrapFreqStab}

\end{document}